\pdfoutput=1 

\documentclass[10pt]{article} 
\usepackage{template}
\usepackage{url,hyperref}
\usepackage[utf8]{inputenc}
\usepackage{mathtools}
\usepackage{graphicx}
\usepackage{verbatim}
\usepackage{amsmath}
\usepackage{amsfonts}
\usepackage[dvipsnames]{xcolor}
\usepackage{url,hyperref}
\usepackage{amssymb}

\usepackage{authblk}

\usepackage[
backend=biber,
style=numeric,
sorting=none
]{biblatex}

    \title{\includegraphics[scale=0.2]{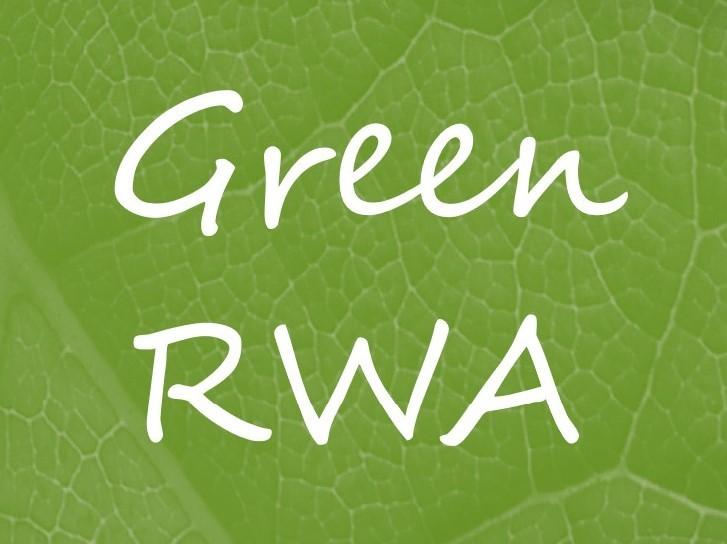}\vfill A Stochastic Climate Model - An approach to calibrate the Climate-Extended Risk Model (CERM)}

    \author[1]{Jean-Baptiste Gaudemet}
    \author[2]{Jules Deschamps}
    \author[3]{Olivier Vinciguerra}
    \affil[1,3]{Green RWA, https://www.greenrwa.org/}
    \affil[2]{Amalthea FS, https://amaltheafs.com/}
    \date{}

\begin{document}

\maketitle

\begin{abstract}

The initial Climate-Extended Risk Model (CERM) \cite{garnier2021climate} addresses the estimate of climate-related financial risk embedded within a bank loan portfolio, through a climatic extension of the Basel II IRB model. It uses a Gaussian copula model calibrated with non stationary macro-correlations in order to reflect the future evolution of climate-related financial risks.

In this complementary article, we propose a stochastic forward-looking methodology to calibrate climate macro-correlation evolution from scientific climate data, for physical and transition efforts specifically. We assume a global physical and transition risk, likened to persistent greenhouse gas (GHG) concentration in the atmosphere. The economic risk is considered stationary and can therefore be calibrated with a backward-looking methodology.
We present 4 key principles to model the GDP and we propose to model the economic, physical and transition effort factors with three interdependent stochastic processes allowing for a calibration with seven precisely defined parameters. These parameters can be calibrated using public data. This new approach means not only to evaluate climate risks without picking any specific scenario but also allows to fill the gap between current one year approach of regulatory and economic capital models and the necessarily long-term view of climate risks by designing a framework to evaluate the resulting credit loss on each step (typically yearly) of the transition path.

This new approach could prove instrumental in the 2022 context of central banks weighing the pros and cons of a climate capital charge.

\vfill


\end{abstract}

\newpage

\section*{Introduction}

Since the very first version of this new climate-related financial risk framework in June 2020 \cite{grwa}, the ecosystem has moved forward a lot. Central Banks have progressed in their analysis of climate change impact on financial stability by running multiple stress-tests exercises. In UK, the Bank of England in 2021 (due in May 2022), French ACPR, ended in 2021. The Network for Greening the Financial System (NGFS) has also grown tremendously, gathering now 105 central banks including the Federal Reserve Bank which joined in January 2021 thanks to the new administration involvement in USA. NGFS refined last year the climate scenari taken in consideration and describes now 6 scenari \cite{ngfs}. Finally, commercial banks have also picked up their games, quickly drafting Net Zero plans and forming coalitions to support the energy transition plan, the latest being 19 banks in North America forming a consortium with the Risk Management Association.

This paper is the result of the continuous forward thinking work Green RWA is fostering, keeping the momentum the CERM has sparkled in its community.
We describe here a calibration methodology allowing to take into account a distribution of scenari (as opposed to scenario picking in the current stress-test approach) through a diffusion model relying on a limited number of well defined parameters. This calibration methodology paves the way for the calculation of credit loss provisions on the net-zero pathway and potentially for a single upfront capital coverage by discounting this curve.
\newline
In the first section, we will introduce four key principles we will use in the second session to define a stochastic diffusion model of climate change. It will allow to derive the CERM macro-correlation parameters, in the third section. As we will see, our calibration methodology is slightly modifying the mathematics of the original CERM as we now need auto-correlated risk factors and a dynamic correlation matrix. We will discuss these model adjustments in the fourth section of the paper. In the fifth section we will study the resulting distribution of the GDP growth curve, which will allow to calculate in the sixth section the asymptotic probability to reach a net-zero economic growth.

The seventh and final section of the paper is a call for a numerical application of the calibration.

\section*{About}

\subsection*{\href{https://www.greenrwa.org/}{Green RWA}}

Green RWA is a unique NGO rooted in the belief that financial institutions have a key role to play in the energy transition. Green RWA supports banks by developing innovative climate financial risk models. It now counts 84 members, bankers, insurers and fintech specialists, and 10 corporates, typically FinTech and Consultancies.

\subsection*{\href{https://amaltheafs.com/}{Amalthea FS}}

Amalthea is a start-up Fintech developing SaaS technology for climate risk and reporting. Amalthea believes banks will need new technology to help achieve the challenges of new regulation and risk frameworks that energy transition and regulation will impose, launching a necessary evolution of their technological infrastructure.

\section*{Acknowledgments}
 
We want to thank specifically Josselin Garnier and Anne Gruz for the high quality interactions we regularly have on the financial climate risk and its modelling. We are also grateful to the central banks and international institutions we met including ECB, FED, European Commission and IMF for their openness and interest in our approach and this calibration methodology. Their expertise and in-depth understanding of the matter was instrumental to narrowing down on this approach.

\newpage

\section{Key principles}
The CERM is calibrated through an estimation of the borrower's assets' sensitivity to the global economic and climate-related financial risks (typically transition and physical risks). In this paper, we suggest to model the joint dynamic of the Global Domestic Product (GDP) between economic cycles, physical climate change damages, and transition costs, with a system of equations derived from the following four key principles:

\begin{itemize}

\item \textbf{P1:} Climate risks are additive to the other economic risks observed historically.
\item \textbf{P2:} Physical risk associated with the global rise of temperature adds up with a growing incertitude over time as the climate reacts to the accumulation in the atmosphere of greenhouse gas (GHG) emissions generated by economic activity \cite{mora2018broad}. This assumption is a general consensus amongst climate specialists, as it is underlying in most climate models.
\item \textbf{P3:}  Accumulated transition efforts reduce future climate change damage by adapting the economy and limiting new greenhouse gas emissions. This typically is represented by net-zero emission plans, aggregated both at state and corporate levels.
\item \textbf{P4:} Transition efforts stem from the political pressure and ecological awareness generated by climate change. Extreme weather events, flooding and large fires around the world typically encourage citizens to accept increasing transition efforts and urge decision-makers to enforce climate adaptation investments and GHG mitigation policies.
\end{itemize}

\section{Modeling climate change}

\subsection{GDP model}

We define a yearly time axis $t\in\mathbb{Z}$. We choose to set $t=0$ as the current year and $t=-t_0$ as the year from which GHG emissions have started to modify climate. We will model the growth of the economy from time $-t_0$ to time $t$ as the additive growth of the log GDP during this climate change period. We introduce three stochastic cumulative growth factors from $-t_0$ to $t$:

\begin{itemize}
    \item $\tilde{Y}_E^t$, the hypothetical climate-free growth of the economy (i.e., if climate change did not exist) since $-t_0$. 
    \item $\tilde{Y}_P^t$, the cumulative physical damage to the economy due to climate change since $-t_0$;
    \item $\tilde{Y}_T^t$, the cumulative economic cost due to transition efforts since $-t_0$.
\end{itemize}

By definition of $-t_0$ we have the initial conditions:  $\tilde{Y}_E^{-t_0}=\tilde{Y}_P^{-t_0}=\tilde{Y}_T^{-t_0}=0$. As per the first principle $\textbf{P1}$, those three random variables add up to model the growth of the economy. 

\begin{equation}
\forall t\in\mathbb{Z}, t\geq -t_0, GDP^{t}=GDP^{-t_0}\exp{\left[\tilde{Y}_E^t-\tilde{Y}_T^t- \tilde{Y}_P^t\right]}
\end{equation}

The variables $\tilde{Y}_E^t$, $\tilde{Y}_P^t$, $\tilde{Y}_T^t$ are fixed for all past (and present) dates $t \in \{-t_0, ..., 0\}$. In the next section we will introduce a forward-looking stochastic diffusion model for $t>0$.

\subsection{Forward-looking climate factors model}

In this section we propose to model the dynamic of the three stochastic growth factors defined above. Since we model the log GDP, we have a year-on-year additive model:

\begin{equation}
\forall t\in\mathbb{N}, 
\begin{cases}
\tilde{Y}_E^{t+1}=\tilde{Y}_E^{t}+R+e\epsilon_E^{t+1} \,\,(\tilde{L}_1) \\
\tilde{Y}_P^{t+1}=\tilde{Y}_P^t+\tilde{\gamma} (\tilde{Y}_E^{t+1}-\tilde{Y}_P^{t+1}-\tilde{Y}_T^{t+1})-\tilde{\alpha} \tilde{Y}_T^{t+1}+\tilde{p}W_P^{t+1} \,\,(\tilde{L}_2) \\
\tilde{Y}_T^{t+1}=\tilde{Y}_T^t+\beta (\tilde{Y}_P^{t}-\tilde{Y}_P^{t-1}) + \theta \epsilon_{\theta}^{t+1}   \,\,(\tilde{L}_3) 
\end{cases},
\end{equation} 
where $R$, $e$, $\tilde{p}$, $\theta$, $\tilde{\alpha}$, $\beta$ $\tilde{\gamma}$ are all positive coefficients to be calibrated, and all $(\epsilon_r^t)_{r \in \{E,P,\theta\}} \sim \mathcal{N}(0,1)$ are i.i.d.\newline

As per the IRB regulatory principle, we first model in equation $(\tilde{L}_1)$ a stationary climate-free economic dynamic $\tilde{Y}_E$ with a stochastic variable $e\epsilon_E^t$ of constant variance, to which we add a constant growth factor R. Note it is possible without changing the formalism of the model to use a deterministic growth factor $R^t$.  \newline

In the second equation $(\tilde{L}_2)$, we increase the physical damage $\tilde{Y}_P$ by the incremental impact on climate change $\tilde{\gamma}(\tilde{Y}_E-\tilde{Y}_P-\tilde{Y}_T)$ resulting from marginal GHG emissions generated by the economy (as per the second principle $\textbf{P2}$). To it is subtracted the benefit $\tilde{\alpha}\tilde{Y}_T$ of the cumulative transition effort (as per the third principle $\textbf{P3}$). This allows to see $\tilde{\gamma}$ as the climate change intensity of the pre-transition economic activity, and $\tilde{\alpha}$ as the transition efficiency. In addition, we integrate the growing climate change incertitude by adding the independent Wiener process $\tilde{p} W_P$. Note it possible without changing the formalism of the model to add an acceleration factor to the climate damage $\tilde{Y}_P^t$ in order to capture secondary climate change effects. \newline

The third equation $(\tilde{L}_3)$ formalizes the forth principle $\textbf{P4}$ by a linear regression between the incremental physical damage to the economy and the additional transition effort made by the economy. We note $\beta$ the intercept term (transition efforts reactivity) and $\theta$ the error term (idiosyncratic transition efforts).\newline

The next section will derive the CERM parameters from the forward-looking GDP model presented.

\subsection{Glossary}

\begin{center}
\begin{tabular}{ |c|c| } 
 \hline
 variable & interpretation \\
 \hline
 $\tilde{Y}_E$ & cumulative hypothetical climate-free economic growth\\
 $\tilde{Y}_P$ & cumulative economic cost due to climate change\\
 $\tilde{Y}_T$ & cumulative economic cost due to transition efforts\\
 $\tilde{Y}_E - \tilde{Y}_P - \tilde{Y}_T$ & cumulative economic growth\\

 \hline
\end{tabular}
\end{center}

\begin{center}
    \begin{tabular}{|c|c|}
    \hline
    parameter & interpretation  \\
    \hline
    $\tilde{\gamma}$ & climate change intensity of the economic activity\\
    $\tilde{p}$ & idiosyncratic physical risk (standard deviation rate) \\
    $\tilde{\alpha}$ & transition efficiency \\ 
    $\beta$ & transition effort reactivity to climate change \\ 
    $\theta$ & independent transition effort (standard deviation)\\ 
    
     $R$ & hypothetical climate-free average growth rate \\ 
     $e$ & idiosyncratic economic risk (standard deviation) \\ 
   
     \hline
    \end{tabular}
\end{center}

\section{CERM parameters}

\subsection{Risk model}

We introduce the risk factors $Y_E^t$, $Y_P^t$, $Y_T^t$, as the random parts of the incremental growth factors of the log GDP between time $t-1$ and $t$. We have: 

\begin{equation}
\forall t \in \mathbb{N},
\begin{cases}
  Y_E^{t+1}=\tilde{Y}_E^{t+1}-\tilde{Y}_E^{t}-\mathbb{E}(\tilde{Y}_E^{t+1}-
  \tilde{Y}_E^{t}) \\
  Y_P^{t+1}=\tilde{Y}_P^{t+1}-\tilde{Y}_P^{t}-\mathbb{E}(\tilde{Y}_P^{t+1}-
  \tilde{Y}_P^{t})\\
  Y_T^{t+1}=\tilde{Y}_T^{t+1}-\tilde{Y}_T^{t}-\mathbb{E}(\tilde{Y}_T^{t+1}-
  \tilde{Y}_T^{t})
\end{cases},
\end{equation} 
with $Y_E^{t}= Y_P^{t}= Y_T^{t}=0$ for $t \leq 0$ since historical growth factors are known. By construction, for $t \geq 0$ all $(Y_r^t)_{r \in \{E,P,\theta\}}$ are centered.

\subsection{Risk covariance}

In order to calculate the correlation of the CERM economic and climate factors, we must derive from (2) the covariance matrix of the risk factors $Y_E$, $Y_P$ and $Y_T$. The difference in equation $(2)$ between times $t+1$ and $t$ reads:

$$
\forall t\in\mathbb{N},
\begin{cases} Y_E^{t+1}=e \epsilon_E^{t+1} \\
(1+\tilde{\gamma})Y_P^{t+1}=Y_P^t+\tilde{\gamma} Y_E^{t+1}-(\tilde{\alpha}+\tilde{\gamma}) Y_T^{t+1} +\tilde{p}\epsilon_P^{t+1}  \\
Y_T^{t+1}=\beta Y_P^t + \theta \epsilon_\theta^{t+1}   \\ 
\end{cases},
$$
with $\epsilon_P^{t+1}=W_P^{t+1}-W_P^{t} \sim \mathcal{N}(0,1)$ (by definition of a Wiener process). Then, by substituting $Y_E^{t+1}$ and $Y_T^{t+1}$ from the first and the third equations in the second one, we get:

$$
\forall t\in\mathbb{N},
\begin{cases} Y_E^{t+1}=e\epsilon_E^{t+1}\\ (1+\tilde{\gamma})Y_P^{t+1}=(1-(\tilde{\alpha}+\tilde{\gamma})\beta)Y_P^t-(\tilde{\alpha}+\tilde{\gamma})\theta\epsilon_{\theta}^{t+1}+ e\tilde{\gamma}\epsilon_E^{t+1}+\tilde{p}\epsilon_P^{t+1} \\
Y_T^{t+1}=\beta Y_P^t + \theta \epsilon_{\theta}^{t+1}
\end{cases}
$$

To simplify the system we introduce the reduced parameters: 
\begin{equation}
\begin{cases}
\alpha=\frac{\tilde{\alpha}}{1+\tilde{\gamma}}, \, \,
\gamma= \frac{\tilde{\gamma}}{1+\tilde{\gamma}}, \, \,
p=\frac{\tilde{p}}{1+\tilde{\gamma}} \\
q=\frac{1-(\tilde{\alpha}+\tilde{\gamma})\beta}{1+\tilde{\gamma}} = 1-\alpha \beta - (1+\beta) \gamma = (1-\gamma) - (\alpha + \gamma)\beta \, \, \text{(is assumed to be $<1$)}.
\end{cases}
\end{equation}

Substituting gives the following equations:
\begin{equation}
\forall t\in\mathbb{N},
\begin{cases} Y_E^{t+1}=e\epsilon_E^{t+1} \, (L_1) \\
Y_P^{t+1}=q Y_P^t-(\alpha+\gamma)\theta\epsilon_{\theta}^{+1}+ \gamma e \epsilon_E^{t+1} +p\epsilon_P^{t+1}\, (L_2)\\
Y_T^{t+1}=\beta Y_P^t + \theta \epsilon_{\theta}^{t+1} \,(L_3) \end{cases}.
\end{equation}

For $t \geq 0$, writing the vector $\mathbf{Y}^t=(Y_E^t \,\, Y_P^t \,\, Y_T^t)^T$ gives:

$$
\forall t\in\mathbb{N}, \, \mathbf{Y^{t+1}}=
\left(\begin{array}{ccc}
0 & 0 & 0 \\
0 & q & 0\\
0 & \beta & 0
\end{array}\right)
\mathbf{Y^t}
+
\left(\begin{array}{c}
e\epsilon_E^{t+1} \\
-(\alpha + \gamma)\theta\epsilon_{\theta}^{t+1}+e\gamma\epsilon_E^{t+1}+p\epsilon_P^{t+1}\\
\theta \epsilon_{\theta}^{t+1}
\end{array}\right).
$$

For the following, we will note:
$$
\mathbf{A}=
\left(\begin{array}{ccc}
0 & 0 & 0 \\
0 & q & 0\\
0 & \beta & 0
\end{array}\right),\,
\; \text{and: } \forall t>0,\, 
\mathbf{E_{t}}=\left(\begin{array}{c}
e\epsilon_E^{t} \\
-(\alpha + \gamma)\theta\epsilon_{\theta}^{t}+e\gamma\epsilon_E^{t}+p\epsilon_P^{t}\\
\theta \epsilon_{\theta}^{t}
\end{array}\right).
$$

By induction (initiated with $\mathbf{Y}^0=0$): $ \forall t \geq 1, \mathbf{Y^t}= \sum_{k=0}^{t-1}\mathbf{A}^k\mathbf{E_{t-k}} \text{, with } \mathbf{A}^0=\mathbf{I_3}$. \newline

As all $(\mathbf{E_k})_k$ are independent, the variance is: 

\begin{equation}
\forall t \geq 1, \, \mathbb{V}[\mathbf{Y^t}]=\mathbb{V}[\sum_{k=0}^{t-1}\mathbf{A}^k \mathbf{E_{t-k}}]=\sum_{k=0}^{t-1} \mathbb{V}[\mathbf{A}^k \mathbf{E_{t-k}}]=\sum_{k=0}^{t-1} \mathbf{A}^k \mathbf{V}(\mathbf{A}^k)^T,
\end{equation}
with

\begin{equation} \mathbf{V}=\mathbb{V}[\mathbf{E_t}]=
\left(\begin{array}{ccc}
e^2 & \gamma e^2 & 0 \\
\gamma e^2 & (\alpha+\gamma)^2\theta^2+e^2\gamma^2 + p^2 & -(\alpha+\gamma)\theta^2\\
0 & -(\alpha+\gamma)\theta^2 & \theta^2
\end{array}\right).
\end{equation}

We introduce the characteristic standard deviation $\sigma=\sqrt{(\alpha+\gamma)^2\theta^2+e^2\gamma^2+ p^2}$ to simplify future calculations. Having $\left| q \right| < 1$ also allows to introduce the sum $c_t=\sum_{k=0}^{t-1}q^{2k}=\frac{1-q^{2t}}{1-q^2}$ for $t \geq 1$ with $c_0=0$.
\newline 

We know by induction that:
\begin{equation}
    \forall t \geq 1,\, \mathbf{A^t}=
\left(\begin{array}{ccc}
0 & 0 & 0 \\
0 & q^t & 0\\
0 & \beta q^{t-1} & 0
\end{array}\right),
\end{equation} and from (6) and (7), we draw:

\begin{equation}
\forall t \geq 1,\, \mathbb{V}[\mathbf{Y^t}]=
\left(\begin{array}{ccc}
e^2 & \gamma e^2 & 0 \\
\gamma e^2 & \sigma^2c_t & -(\alpha + \gamma)\theta^2+\sigma^2 \beta qc_{t-1} \\
0 &-(\alpha + \gamma)\theta^2+\sigma^2\beta q c_{t-1}  & \theta^2+\beta^2\sigma^2c_{t-1}
\end{array}\right)
\end{equation}

\subsection{CERM macro-correlations}

In the CERM, the macro-correlations at time $t+1$ are defined as the respective sensitivities to the economic and climate risk factors between time $t$ and $t+1$. We can now derive these quantities as the standard deviations of the incremental growth factors:

\begin{equation}
\forall t\in\mathbb{N+},
\begin{cases}
  \xi_E^{t}=\sqrt{\mathbb{V}[Y_E^{t}]}=e \\
  \xi_P^{t}=\sqrt{\mathbb{V}[Y_P^{t}]}=\sigma\sqrt c_t\\
  \xi_T^{t}=\sqrt{\mathbb{V}[Y_T^{t}]}=\sqrt{\theta^2+\beta^2\sigma^2c_{t-1}}
\end{cases}
\end{equation}

We can now calculate the correlation matrix for CERM risk factors $\mathbf{Z^t}\sim \mathcal{N}(0,C_t)$ at time $t \geq 1$ as: $C_t=Corr[Y_E^{t}, -Y_P^{t}, -Y_T^{t}]$. Using equations (7) and (9), we get:
    
\begin{equation}
C_t=
\left(\begin{array}{ccc}
1 & -\frac{\gamma e^2}{\xi_E^t \xi_P^t} & 0 \\
-\frac{\gamma e^2}{\xi_E^t \xi_P^t} & 1 & \frac{\sigma^2\beta qc_{t-1}-(\alpha+\gamma)\theta^2}{\xi_P^t\xi_T^t} \\\
0 & \frac{\sigma^2\beta qc_{t-1}-(\alpha+\gamma)\theta^2}{\xi_P^t\xi_T^t} & 1
\end{array}\right)
\end{equation}

It is noteworthy that the current version of the CERM cannot integrate time-dependent correlation matrices. In the next section we will see how to adapt the CERM accordingly.

\subsubsection{Asymptotic study}

The macro-correlations of equation (8) converge to: 
    
\begin{equation}
    \begin{cases} 
\xi_E^\infty=e \\
\xi_P^\infty=\sigma/\sqrt{1-q^2} \\ \xi_T^\infty=\sqrt{\frac{\beta^2\sigma^2}{1-q^2}+\theta^2}\
\end{cases},
\end{equation}
from which we can calculate the limit $C_{\infty}$ of the correlation matrix. With $c = \beta^2 + \frac{\theta^2 (1-q^2)}{\sigma^2}$, we have:

\begin{equation}
C_{\infty}=
\left(\begin{array}{ccc}
1 & -\frac{\gamma e \sqrt{1-q^2}}{\sigma} & 0 \\
-\frac{\gamma e \sqrt{1-q^2}}{\sigma} & 1 & \frac{(1-\gamma)\beta}{\sqrt{c}}-(\alpha + \gamma) \sqrt{c} \\
0 & \frac{(1-\gamma)\beta}{\sqrt{c}}-(\alpha + \gamma) \sqrt{c} & 1
\end{array}\right)
\end{equation} 

As expected, we note that the asymptotic correlation between physical and transition risk decreases with the idiosyncratic transition effort parameter $\theta$ and is negative in the absence of reactivity of transition efforts to climate change ($\beta=0$).

\subsection{Auto-correlations}

In this section we analyze how the risk factors previously presented are auto-correlated. We introduce the delay $\tau>0$, giving for $t \geq 0$: $\mathbf{Y^{t+\tau}}=A^{\tau}\mathbf{Y^t}+\sum_{k=0}^{\tau-1}A^k \mathbf{E_{t+\tau-k}}$. Having this formula for $\mathbf{Y^{t+\tau}}$, we can derive:

\begin{equation}
    Cov[\mathbf{Y^{t+\tau}},\mathbf{Y^t}]=A^{\tau}\mathbb{E}[\mathbf{Y^t}\mathbf{Y^t}^T]+\sum_{k=0}^{\tau-1}A^k \mathbb{E}[\mathbf{E_{t+\tau-k}}\mathbf{Y^t}^T]=A^{\tau}\mathbb{E}[\mathbf{Y^t}\mathbf{Y^t}^T]=A^{\tau} \mathbb{V}[\mathbf{Y^t}]
\end{equation}
because $\mathbf{Y^t}$ is centered and independent from $\mathbf{E_{t+\tau-k}}$ for $k \in \{0,...,\tau-1\}$.\newline

Here, we have:
\begin{equation}
    Cov[\mathbf{Y^{t+\tau}},\mathbf{Y^t}] =
\left(\begin{array}{ccc}
0 & 0 & 0 \\
\gamma e^2 q^{\tau} & \sigma^2 q^{\tau} c_t & -(\alpha + \gamma)\theta^2 q^\tau + \sigma^2 \beta q^{\tau+1}c_{t-1} \\
\beta \gamma e^2 q^{\tau-1} & \sigma^2 \beta q^{\tau-1} c_t & -(\alpha + \gamma)\beta\theta^2 q^{\tau-1} + \sigma^2 \beta^2 q^{\tau}c_{t-1}
\end{array}\right),
\end{equation}
giving:
$$Cov[\mathbf{Y^{t+\tau}},\mathbf{Y^t}] \xrightarrow{t \rightarrow \infty}
\left(\begin{array}{ccc}
0 & 0 & 0 \\
\gamma e^2 q^{\tau} & \frac{\sigma^2 q^{\tau}}{1-q^2} & -(\alpha + \gamma)\theta^2 q^\tau + \frac{\sigma^2 \beta q^{\tau+1}}{1-q^2} \\
\beta \gamma e^2 q^{\tau-1} & \frac{\sigma^2 \beta q^{\tau-1}}{1-q^2} & -(\alpha + \gamma)\beta\theta^2 q^{\tau-1} + \frac{\sigma^2 \beta^2 q^{\tau}}{1-q^2}
\end{array}\right).$$

And then, as $\boldsymbol\xi^t$ is the standard deviation vector of $\mathbf{Y^t}$ (meaning, the square root of the diagonal of its co-variance matrix), we note $\frac{1}{\boldsymbol\xi^t}$ the vector whose coefficient are the inverses of the coefficients of $\boldsymbol\xi^t$. We have:
$$\frac{1}{\boldsymbol\xi^t} \left(\frac{1}{\boldsymbol\xi^{t+\tau}}\right)^T = 
\begin{matrix}
((\frac{1}{\sigma(\mathbf{Y^t})_i\sigma(\mathbf{Y^{t+\tau}})_j})_{i,j})
\end{matrix}
$$
and then ($*$ is coefficient-by-coefficient product):

\begin{equation}
    Corr(\mathbf{Y^{t+\tau}},\mathbf{Y^t})=\frac{1}{\boldsymbol\xi^t} \left(\frac{1}{\boldsymbol\xi^{t+\tau}}\right)^T * Cov[\mathbf{Y^{t+\tau}},\mathbf{Y^t}]. 
\end{equation}

From this we can draw:

\begin{equation}
    \lim\limits_{t \to \infty} Corr(\mathbf{Y^{t+\tau}},\mathbf{Y^t}) = \left(\begin{array}{ccc}
    0 & 0 & 0 \\
    \frac{\gamma e^2 q^{\tau} \sqrt{1-q^2}}{\sigma} & q^{\tau} & \frac{\beta q^{t+1}}{\sqrt{c}} - \frac{(\alpha + \gamma) \theta^2 q^{\tau} (1-q^2)}{\sigma^2 \sqrt{c}} \\
    \frac{\beta \gamma e q^{\tau-1} \sqrt{1-q^2}}{\sigma \sqrt{c}} & \frac{\beta q^{\tau-1}}{\sqrt{c}} & \frac{\beta^2 q^{\tau}}{c}-\frac{\beta(\alpha+\gamma)\theta^2 q^{\tau-1} (1-q^2)}{\sigma^2 c}.
\end{array}\right)
\end{equation}

We notice that physical and transition risks are auto- and cross-correlated. It is quite a fundamental difference with the original CERM as it was considering risk factors as independently distributed. We believe this model better reflects the reality of climate change which is obviously a persistent phenomenon. The economic risk stays without auto-correlation in order to model punctual economic crisis consistently with the regulatory IRB model. 

\section{Adaptation of the CERM}

The following approach is based on the third approach developed in the original CERM \cite{garnier2021climate}. For coherence purposes, it adopts the same template of presentation. As mentioned previously, the time-dependent model for auto-correlations calls for an adaptation of the original CERM paper.\newline

In the following, as is in the original CERM paper, $\alpha_{g,i,t,j}$ are the micro-correlation adjustment parameters (each borrower in group $g$ with rating $i$ at time $t$ has a micro-correlation $\alpha_{g,i,t,j}$ with risk $j$), and $\xi_{t,j}$ are the macro-correlation parameters, as presented earlier on. The correlation coefficient $R_{g,i,t}$ is the proportion of the variance  of the normalized log asset value that is due to systematic risks. The corresponding loading factors are noted $a_{g,i,t}$. Also, $M_{g,t}$ is the migration matrix for borrowers in group $g$ at time $t$: a borrower in group $g$ with rating $i$ at time $t$ will move to a new rating $j$ with corresponding probability $M_{g,t,i,j}$. The $z_{g,t,i,j}$ coefficients are the corresponding thresholds: the same borrower will move to that same new rating if their log asset value falls behind it. \newline

Like in the original CERM model, we decide here that:
\begin{itemize}
    \item the time unit is one year (horizon of the migration matrices);
    \item at time 1 the migration matrix $M_{g,1}$ is equal to $M_g^{reg}$ and the correlation $R_{g,i,1}$ is determined by the regulator's formula;
    \item the migration matrices and the regulator’s formula for the correlation are updated at time $t \geq 2$ because, contrary to the economic and idiosyncratic risks, which are stationary, the physical and transition efforts evolve with time;
    \item the factor loadings $a_{g,i,t,j}$ are proportional to the product of the macro-correlation and micro-correlation adjustment parameters.
\end{itemize}

We note:
$$M_{g,1}=M_g^{reg},$$
$$R_{g,i,1}=R_{g,i}^{reg},\,\,R_{g,i}^{reg}=\mathbf{R}((M_g^{reg})_{iK}) \text{ (the } \mathbf{R} \text{ correlation function is given in the original paper)},$$
$$a_{g,i,1}=a_{g,i}^{reg}=\sqrt{R_{g,i}^{reg}}\frac{\tilde{a}_{g,i,1}}{\sqrt{\tilde{a}_{g,i,1} \cdot C_1 \tilde{a}_{g,i,1}}}$$

Similarly to the approach 3, we define (all products and quotients are coefficient-by-coefficient) $$\tilde{a}_{g,i,t}=\alpha_{g,i,t}*\xi_{t},\,\, c_{g,i,t}=\frac{a_{g,i,1}*\tilde{a}_{g,i,t}}{\tilde{a}_{g,i,1}}.$$

At time $t \geq 1$, we have:
$$(M_{g,t})_{ij}=
\begin{cases}
1-\Phi(z_{g,t,i2}) & j=1,\\
\Phi(z_{g,t,ij})-\Phi(z_{g,t,ij+1}) & 2\leq j \leq K-1, \\
\Phi(z_{g,t,iK}) & j=K,
\end{cases}$$
with
$$z_{g,t,ij}=\frac{z_{g,ij}^{reg}}{\sqrt{1+c_{g,i,t} \cdot C_t c_{g,i,t}-a_{g,i}^{reg} \cdot C_1 a_{g,i}^{reg}}},$$
$$z_{g,ij}^{reg}=\Phi^{-1}(\sum_{j'=j}^K (M_g^{reg})_{ij'}),$$
and we also have
$$R_{g,i,t}=\frac{c_{g,i,t} \cdot C_t c_{g,i,t}}{1+c_{g,i,t} \cdot C_t c_{g,i,t}-a_{g,i}^{reg} \cdot C_1 a_{g,i}^{reg}}$$
$$a_{g,i,t}=\frac{c_{g,i,t}}{\sqrt{1+c_{g,i,t} \cdot C_t c_{g,i,t}-a_{g,i}^{reg} \cdot C_1 a_{g,i}^{reg}}}$$

Both formulas for the loading factors and for the correlations coincide, as proven hereunder.

\subsection{Proof}

At time 1, the normalized log asset value is given by
$$X_1^{(q)}=a_{g,i}^{reg} \cdot Z_1 + \sqrt{1-a_{g,i}^{reg} \cdot C_1 a_{g,i}^{reg}} \epsilon_1^{(q)}.$$

In this approach, the idiosyncratic risk is stationary while the micro-correlation and macro-correlation parameters evolve in time. This means that we have in fact
$$\bar{X}_t^{(q)}=c_{g,i,t} \cdot Z_t + \sqrt{1-a_{g,i}^{reg} \cdot C_1 a_{g,i}^{reg}} \epsilon_t^{(q)},$$
which is a Gaussian variable with mean zero and variance $1+c_{g,i,t} \cdot C_t c_{g,i,t}-a_{g,i}^{reg} \cdot C_1 a_{g,i}^{reg}.$ As a consequence, the probabilities of rating changes are
$$(M_{g,t})_{ij}=\mathbb{P}(\bar{X}_t^{(q)} \in [z_{g,ij+1}^{reg},z_{g,ij}^{reg}]),$$
where the $z_{g,ij}^{reg}$s are the threshold values associated with the given unconditional matrix $M_g^{reg}$. Furthermore, after normalization, the log asset value (i.e. $X_t^{(q)}=\bar{X}_t^{(q)}/{\sqrt{1+c_{g,i,t} \cdot C_t c_{g,i,t}-a_{g,i}^{reg} \cdot C_1 a_{g,i}^{reg}}}$) has now the form:
$$X_t^{(q)}=a_{g,i,t} \cdot Z_t + \sqrt{1-a_{g,i,t} \cdot C_t a_{g,i,t}} \epsilon_t^{(q)},$$
with $a_{g,i,t}$ defined earlier on. 

This normalized formula which is a function of the risk factor correlation $C_t$ at time t only, results from the following calculation. Using the definition of the correlation with systemic risk factors:
$$R_{g,i,t}=\frac{c_{g,i,t} \cdot C_t c_{g,i,t}}{1+c_{g,i,t} \cdot C_t c_{g,i,t}-a_{g,i}^{reg} \cdot C_1 a_{g,i}^{reg}},$$
we get
$$\sqrt{1+c_{g,i,t} \cdot C_t c_{g,i,t}-a_{g,i}^{reg} \cdot C_1 a_{g,i}^{reg}}=\sqrt{\frac{1-R_{g,i}^{reg}}{1-R_{g,i,t}}}$$
and that is why
$$X_t^{(q)}=\frac{\bar{X}_t^{(q)}}{\sqrt{1+c_{g,i,t} \cdot C_t c_{g,i,t}-a_{g,i}^{reg} \cdot C_1 a_{g,i}^{reg}}}=\frac{c_{g,i,t} \cdot Z_t + \sqrt{1-a_{g,i}^{reg} \cdot C_1 a_{g,i}^{reg}} \epsilon_t^{(q)}}{\sqrt{1+c_{g,i,t} \cdot C_t c_{g,i,t}-a_{g,i}^{reg} \cdot C_1 a_{g,i}^{reg}}}$$
$$=a_{g,i,t} \cdot Z_t + \frac{\sqrt{1-R_{g,i}^{reg}}}{\sqrt{\frac{1-R_{g,i}^{reg}}{1-R_{g,i,t}}}}\epsilon_t^{(q)}=a_{g,i,t} \cdot Z_t + \sqrt{1-R_{g,i,t}}\epsilon_t^{(q)}$$
$$=a_{g,i,t} \cdot Z_t + \sqrt{1-a_{g,i,t} \cdot C_t a_{g,i,t}}\epsilon_t^{(q)}$$

\section{The GDP distribution}

In this section, we want to calculate the distribution of the GDP as a function of time. The equation (1) models the GDP as a stochastic process $GDP^t$ following at each time t a log-normal distribution law of parameters: $$\mu^t=\mathbb{E}(\tilde{Y}_E^t-\tilde{Y}_P^t-\tilde{Y}_T^t) \text{ and }  (s^t)^2=\mathbb{V}(\tilde{Y}_E^t-\tilde{Y}_P^t-\tilde{Y}_T^t).$$ 

As per the classic log-normal distribution formulas, we therefore have at each time $t$: $$med(GDP^t)=GDP^{-t_0} \exp{(\mu^t)}, \,\,\mathbb{E}({GDP^t})=GDP^{-t_0} \exp(\mu^t+(s^t)^2/2),$$ $$\text{ and } \mathbb{V}({GDP^t})=(GDP^{-t_0})^2 (\exp{((s^t)^2})-1)\exp{(2\mu^t +(s^t)^2)}.$$

\subsection{Expected cumulative risk factors}

With similar calculations as in 3.2, we get:

\begin{equation}
    \forall t\in\mathbb{N},
\begin{cases}
\mathbb{E}(\tilde{Y}_E^{t+1}-\tilde{Y}_E^{t})= R \\
\mathbb{E}(\tilde{Y}_P^{t+1}-\tilde{Y}_P^{t})=q\mathbb{E}(\tilde{Y}_P^{t}-\tilde{Y}_P^{t-1}) + \gamma R \\
\mathbb{E}(\tilde{Y}_T^{t+1}-\tilde{Y}_T^{t})=\beta \mathbb{E}(\tilde{Y}_P^{t}-\tilde{Y}_P^{t-1})  \ 
\end{cases}
\end{equation}

The second equation is that of an arithmetico-geometric sequence. Since $\tilde{\mathbf{Y}}^k$ are known for $k \leq 0$, this gives:

\begin{equation}
\forall t\in\mathbb{N},
\begin{cases}
\mathbb{E}(\tilde{Y}_E^{t+1})=\tilde{Y}_E^0+(t+1) R\\
\mathbb{E}(\tilde{Y}_P^{t+1})=\tilde{Y}_P^0+\frac{\gamma R}{1-q}(t+1)+\left((\tilde{Y}_P^0-\tilde{Y}_P^{-1}) - \frac{\gamma R}{1-q} \right)q\frac{1-q^{t+1}}{1-q}\\
\mathbb{E}(\tilde{Y}_T^{t+1}) = \tilde{Y}_T^0 + \beta (\tilde{Y}_P^0-\tilde{Y}_P^{-1}) + \beta \left(\frac{\gamma R}{1-q}t+\left((\tilde{Y}_P^0-\tilde{Y}_P^{-1}) - \frac{\gamma R}{1-q} \right) q \frac{1-q^t}{1-q} \right) \ 
\end{cases}
\end{equation}

\subsection{Asymptotic study}

\subsubsection{Log-GDP asymptotic median growth}

As a result, the median growth of the economy is asymptotically linear:

\begin{equation}
\mu^t = r_{\mu}^\infty t  + \mu_H + o (1),
\end{equation}

with 

\begin{equation}
r_{\mu}^\infty=\frac{\alpha \beta R}{\alpha \beta + (1+\beta)\gamma}
\end{equation}

and

\begin{equation}
    \mu_H = \mu^0 + \left((\tilde{Y}_P^0 - \tilde{Y}_P^{-1})-\frac{\gamma R}{1-q} \right) \times \left[1-\frac{1-\beta}{\alpha \beta + (1+\beta)\gamma} \right]
\end{equation}

From this, we draw the conclusion that although the climate-related factors cannot reverse the median economic growth, they do asymptotically curb it, as $\frac{\alpha \beta R}{\alpha \beta + (1+\beta)\gamma} \leq R$. The more efficient the transition is in mitigating and adapting the economy to carbon emissions (parameter $\alpha$, in comparison with $\gamma$), the closest the median economic growth is to the climate-free growth ($R$). 

\subsubsection{Log-GDP asymptotic variance}

We introduce the vector $\mathbf{\tilde{Y}}=(\tilde{Y}_E \,\, \tilde{Y}_P \,\, \tilde{Y}_T)^T$. We first want to calculate $\mathbb{V}[\mathbf{\tilde{Y}}^t]$ for $t \geq 0$. $\mathbf{\tilde{Y}}^0$ being known (as in constant), $\mathbb{V}[\mathbf{\tilde{Y}}^t]=\mathbb{V}[\mathbf{\tilde{Y}}^T-\mathbf{\tilde{Y}}^0]$. From equation (3) we derive: 

$$\mathbf{\tilde{Y}}^{t}-\mathbf{\tilde{Y}}^0=\sum_{k=1}^{t}\mathbf{Y^k}+C=\sum_{k=1}^{t}\sum_{i=0}^{k-1}\mathbf{A}^i\mathbf{E_{k-i}}+C  \textrm{ with } \mathbf{A}^0=\mathbf{id} \textrm{ and where } C \textrm{ is a constant.}$$

From this we draw:

$$\mathbb{V}[\mathbf{\tilde{Y}^t}] = \mathbb{V}\left[\sum_{k=1}^{t}\mathbf{Y^k}\right]=\mathbb{V}\left[\sum_{k=1}^{t}\sum_{i=0}^{k-1}\mathbf{A}^i \mathbf{E_{k-i}}\right]$$

With a classic manipulation of the double sum, we get:

\begin{equation}
\mathbb{V}\left[\sum_{k=1}^{t}\mathbf{Y^k}\right]=\mathbb{V}\left[\sum_{k=1}^{t} \left(\sum_{i=0}^{t-k} \mathbf{A}^i \right)  \mathbf{E_k}\right]
\end{equation}

And since the random vectors $\mathbf{E_k}$ are i.i.d., we have:
\begin{equation}
\mathbb{V}[\mathbf{\tilde{Y}^t}]=\sum_{k=1}^{t}\mathbb{V}\left[ \left(\sum_{i=0}^{t-k}\mathbf{A}^i \right) \mathbf{E_k}\right]=\sum_{k=1}^{t} \left[ \left(\sum_{i=0}^{t-k}\mathbf{A}^i \right) \mathbb{V}(\mathbf{E_k}) \left(\sum_{i=0}^{t-k}\mathbf{A}^i \right)^T \right]=\sum_{k=1}^{t} \left[ \left(\sum_{i=0}^{t-k}\mathbf{A}^i \right) \mathbf{V}\left(\sum_{i=0}^{t-k}\mathbf{A}^i \right)^T \right] 
\end{equation} using the same notations as $(6)$, which gives in turn:

\begin{equation}
\mathbb{V}[\mathbf{\tilde{Y}^t}]=\sum_{k=1}^{t} \left[ \left(\sum_{i=0}^{k-1}\left[\mathbf{A}^i\right] \right) \mathbf{V} \left(\sum_{i=0}^{k-1}\left[(\mathbf{A}^i)\right] \right)^T \right].
\end{equation}

Introducing the sum $ b_k=\sum_{i=0}^{k-1} q^i = \frac{1-q^k}{1-q}$ for $t \geq 1$ with $s_0=0$, we get:

$$
\forall k \in\mathbb{N^*},\,\sum_{i=0}^{k-1}\left[\mathbf{A^i}\right]=
\left(\begin{array}{ccc}
1 & 0 & 0 \\
0 & b_k & 0\\
0 & \beta b_{k-1} & 1
\end{array}\right).
$$

Hence the covariance matrix of the GDP is:

\begin{equation}
\mathbb{V}[\mathbf{\tilde{Y}}^t]=\sum_{k=1}^{t}\left(\begin{array}{ccc}
e^2 & \gamma e^2 b_k & \beta\gamma e^2 b_{k-1} \\
\gamma e^2 b_k & \sigma^2 b_k^2 & \beta\sigma^2 b_k b_{k-1}-(\alpha+\gamma)\theta^2 b_k\\\
\beta\gamma e^2 b_{k-1}&\beta\sigma^2 b_k b_{k-1}-(\alpha+\gamma)\theta^2 b_k & \theta^2+\beta^2\sigma^2 b_{k-1}^2-2\beta\alpha\theta^2 b_{k-1}
\end{array}\right).
\end{equation}

This allows to compute $s^t$ and $\mathbb{V}[GDP^t]$, as ${(s^t)}^2 = \mathbb{V}[\tilde{Y}_E^t-\tilde{Y}_P^t-\tilde{Y}_T] = (1\,-1\,-1) \mathbb{V}[\mathbf{\tilde{Y}}^t]\left(\begin{array}{c}
    1  \\
    -1 \\
    -1
\end{array} \right).$

More specifically, asymptotically we have:

\begin{equation}
\left(s^t\right)^2 = r_{s^2}^{\infty} t + s_H^2 + o(1),
\end{equation}
with: \begin{equation}
    r_{s^2}^{\infty} = \left[e^2 + \theta^2 - 2 \frac{\gamma \theta^2 + (1+\beta)(\alpha \theta^2 + \gamma e^2)}{1-q} + \frac{\sigma^2 (1+\beta)^2}{(1-q)^2} \right].
\end{equation}

\subsection{GDP statistics}

Having asymptotic equivalents of $\mu^t$ and ${(s^t)}^2$ draws:

\begin{itemize}
    \item $\log(med(GDP^t)) \sim r_{\mu}^{\infty}t$;
    \item $\log(\mathbb{E}[GDP^t]) \sim \left(\frac{r_{s^2}^{\infty}}{2} + r_{\mu}^{\infty} \right) t$;
    \item $\log(\mathbb{V}[GDP^t]) \sim 2 \left(r_{s^2}^{\infty} + r_{\mu}^{\infty} \right) t.$
\end{itemize}

\section{The probability of a net-zero transition}

The following sections give several distinct approaches to assess the probability of asymptotic net-zero transition, i.e. the event that in the long run physical risks will be impeded enough so that economic development is sustainable.
To that end, we propose 3 approaches, related to the asymptotic behavior of the physical cost factor in regards to the growth factor of the economy. They do not assess the exact same quantities:
\begin{itemize}
    \item the first approach computes the unconditional probability of net-zero transition;
    \item the second one assesses the probability of having no degradation of climate conditionally to the median economic growth;
    \item the last one reflects the probability of having no degradation of climate conditionally to asymptotic positive economic growth.
\end{itemize}

To simplify the following calculations, we introduce some notations to describe the distributions of the incremental growth factors of physical risks and economic development, for $t \in \mathbb{N}$.\newline

We note:
\begin{itemize}
    \item $\tilde{Y}_P^{t+1}- \tilde{Y}_P^t \sim \mathcal{N}(\mu_1^{t+1}, (\sigma_1^{t+1})^2)$;
    \item $(\tilde{Y}_E^{t+1}-\tilde{Y}_P^{t+1}-\tilde{Y}_T^{t+1}) - (\tilde{Y}_E^t-\tilde{Y}_P^t-\tilde{Y}_T^t) \sim \mathcal{N}(\mu_2^{t+1}, (\sigma_2^{t+1})^2);$
    \item $\rho^{t+1} = Corr \left(\tilde{Y}_P^{t+1}- \tilde{Y}_P^t,(\tilde{Y}_E^{t+1}-\tilde{Y}_P^{t+1}-\tilde{Y}_T^{t+1}) - (\tilde{Y}_E^t-\tilde{Y}_P^t-\tilde{Y}_T^t)\right)= Corr \left(Y_P^{t+1}, Y_E^{t+1}-Y_P^{t+1}-Y_T^{t+1} \right),$
\end{itemize}

where:
\begin{itemize}
    \item $\mu_1^{t+1}=\mathbb{E}[\tilde{Y}_P^{t+1}-\tilde{Y}_P^t] = \gamma R b_t + q^t \left(\tilde{Y}_P^0 - \tilde{Y}_P^{-1} \right) \xrightarrow{t \rightarrow \infty} \mu_1 = \frac{\gamma R}{1-q};$
    \item $\mu_2^{t+1}=\mathbb{E}[(\tilde{Y}_E^{t+1}-\tilde{Y}_P^{t+1}-\tilde{Y}_T^{t+1}) - (\tilde{Y}_E^t-\tilde{Y}_P^t-\tilde{Y}_T^t)] = R \left(1-\gamma(b_t + \beta b_{t-1})\right) \xrightarrow{t \rightarrow \infty} \mu_2 = r_{\mu}^{\infty} = \frac{\alpha \beta R}{\alpha \beta + (1+\beta) \gamma};$
    \item  $({\sigma_1^{t+1}})^2=\sigma^2 c_{t+1} \xrightarrow{t \rightarrow \infty} \sigma_1^2 = \frac{\sigma^2}{1-q^2};$
    \item  $({\sigma_2^{t+1}})^2 = (1-2\gamma)e^2 +(1+2\alpha+2\gamma)\theta^2+\sigma^2 \left( c_{t+1} +(\beta^2 - 2\beta q)c_t \right) \xrightarrow{t \rightarrow \infty} \sigma_2^2 = (1-2\gamma)e^2 + (1+ 2\alpha +2\gamma)\theta^2 + \frac{\sigma^2}{1-q^2}\left(1+\beta^2 - 2\beta q\right);$
    \item $\rho^{t+1} = \frac{\sigma^2 c_{t+1} - \gamma e^2 + (\alpha +\gamma)\theta^2 - \sigma^2 \beta q c_t}{\sqrt{\sigma^2 c_{t+1} \times \left((1-2\gamma)e^2 + \theta^2 + 2(\alpha+\gamma)\theta^2+\sigma^2 c_{t+1} +\sigma^2(\beta^2 - 2\beta q)c_t \right)}} \xrightarrow{t \rightarrow \infty} \rho = \frac{1-\beta q + \left((\alpha + \gamma) \theta^2 - \gamma e^2 \right)\frac{1-q^2}{\sigma^2}}{\sqrt{1+\beta^2 - 2\beta q + \frac{1-q^2}{\sigma^2} \left( (1-2\gamma)e^2 + (1+2\alpha +2\gamma) \theta^2 \right)}}.$
    \item $\Phi$ is the cumulative distribution of the standard Gaussian distribution.
    \item $\Phi_2(\cdot\,,\cdot \, ; \rho)$ is the bivariate cumulative Gaussian distribution with correlation $\rho$, as found in \cite{garnier2021climate}.
\end{itemize}

\subsection{Unconditional net zero probability}

We introduce $P_{NZ}^1$ the asymptotic unconditional probability to reach a net-zero transition. It is worthwhile to note that the auto-correlation analysed in section 3.4, guarantees to stay with a high probability close to net-zero for a certain period of time once it has been reached. We define $P_{NZ}^1$ as the limit of the probability to have no degradation of climate, i.e.:
\begin{equation}
    P_{NZ}^1=\lim\limits_{t \to \infty}\mathbb{P} \left[\tilde{Y}_P^{t+1} - \tilde{Y_P^t} < 0 \right].
\end{equation}

Using the  cumulative distribution function of the Gaussian distribution we get: 

\begin{equation}
    P_{NZ}^1 = \Phi \left( -\frac{\gamma R}{\sigma} \sqrt{\frac{1+q}{1-q}} \right) = \Phi \left( -\frac{\gamma R}{\sigma} \sqrt{\frac{2- \alpha \beta - (1+\beta) \gamma}{\alpha \beta + (1+\beta) \gamma}} \right).
\end{equation}

The net-zero probability $P_{NZ}^1$ increases with the transition efficiency $\alpha$ and the transition reactivity $\beta$ which was expected. In addition, $P_{NZ}^1$ is naturally decreasing with the climate-free average economic growth $R$, meaning we cannot grow the economy with total climate impunity. However, we will see in the next two sections that our model makes it possible to reach a net-zero growing economy. 

\subsection{Net zero probability conditional to the median economic growth}

We now define $P_{NZ}^2(r)$ the asymptotic probability to reach a net-zero transition conditionally to the median economic growth analysed in section 5.2.1. Here we note:
\begin{equation}
    P_{NZ}^2(r)=\lim\limits_{t \to \infty}\mathbb{P}\left( \left[\tilde{Y}_P^{t+1} - \tilde{Y_P^t} < 0\, | \, (\tilde{Y}_E^{t+1}-\tilde{Y}_P^{t+1}-\tilde{Y}_T^{t+1}) - (\tilde{Y}_E^t-\tilde{Y}_P^t-\tilde{Y}_T^t) = r \right] \right).
\end{equation}

For $t \in \mathbb{N}$: $$\left(\tilde{Y}_P^{t+1}- \tilde{Y}_P^t \, | \, (\tilde{Y}_E^{t+1}-\tilde{Y}_P^{t+1}-\tilde{Y}_T^{t+1}) - (\tilde{Y}_E^t-\tilde{Y}_P^t-\tilde{Y}_T^t) = r \right)\sim \mathcal{N}\left(\mu_1^{t+1} + \rho^{t+1} \frac{\sigma_1^{t+1}}{\sigma_2^{t+1}}(r-\mu_2^{t+1}),\, (1-(\rho^{t+1})^2) (\sigma_1^{t+1})^2\right).$$

Asymptotically, for $r=\mu_2=r_{\mu}^{\infty}$ (i.e. conditionally to a growth equal to the asymptotic median growth of the economy in the long run):

\begin{equation}
    P_{NZ}^2=\lim_{t \rightarrow \infty}\mathbb{P}\left(\tilde{Y}_P^{t+1}- \tilde{Y}_P^t <0 \, | \, (\tilde{Y}_E^{t+1}-\tilde{Y}_P^{t+1}-\tilde{Y}_T^{t+1}) - (\tilde{Y}_E^t-\tilde{Y}_P^t-\tilde{Y}_T^t) = r_{\infty} \right)= \Phi\left(-\frac{\mu_1}{\sigma_1 \sqrt{1-\rho^2}} \right)
\end{equation}
$$=\Phi \left( -\frac{\gamma R}{1-q}\sqrt{\frac{ \frac{\sigma^2}{1-q^2}(1+\beta^2 - 2\beta q) + (1-2\gamma)e^2 + (1+2\alpha+2\gamma)\theta^2}{(\frac{\sigma^2}{1-q^2})^2 \beta^2 (1+q^2) + \frac{\sigma^2}{1-q^2} [e^2 (1+2\beta q \gamma )+ \theta^2 (-2\beta q (\alpha+\gamma))] - \gamma^2 e^4 + (\alpha+\gamma)^2 \theta^4 - 2(\alpha + \gamma) \theta^2 \gamma e^2}} \right)$$. 

\subsection{Net-zero probability conditional to a positive economic growth}

We now define $P_{NZ}^3$ as the asymptotic probability to reach a net-zero transition conditionally to a positive economic growth:

\begin{equation}
    P_{NZ}^3= \lim\limits_{t \to \infty}\mathbb{P}\left( \left[\tilde{Y}_P^{t+1} - \tilde{Y_P^t} < 0\, | \, (\tilde{Y}_E^{t+1}-\tilde{Y}_P^{t+1}-\tilde{Y}_T^{t+1}) - (\tilde{Y}_E^t-\tilde{Y}_P^t-\tilde{Y}_T^t) >0\right] \right).
\end{equation}

Which gives:

\begin{equation}
    P_{NZ}^3
    = \frac{\Phi (\frac{\mu_1}{\sigma_1}) -\Phi_2(-\frac{\mu_1}{\sigma_1}, -\frac{\mu_2}{\sigma_2}; \rho)}{\Phi (\frac{\mu_2}{\sigma_2})}.
\end{equation}

\subsubsection{Proof}

We consider the random vector $X=(X_1, X_2)^T \sim \mathcal{N}\left((\mu_1, \mu_2)^T, \left(\begin{array}{cc}
    \sigma_1^2 & \rho \sigma_1 \sigma_2 \\
    \rho \sigma_1 \sigma_2 & \sigma_2^2
\end{array}\right)\right)$ with $0<\rho<1$, we have: 
$$\mathbb{P}(X_1<0 \, | \, X_2>0) = \frac{\mathbb{P}(X_1<0 \, \cap \, X_2>0)}{\mathbb{P}(X_2>0)} = \frac{\mathbb{P}(X_1<0)-\mathbb{P}(X_1<0 \, \cap \, X_2 \leq 0)}{\mathbb{P}(X_2>0)} = \frac{\Phi (\frac{\mu_1}{\sigma_1}) -\Phi_2(-\frac{\mu_1}{\sigma_1}, -\frac{\mu_2}{\sigma_2}; \rho)}{\Phi (\frac{\mu_2}{\sigma_2})}.$$

\section{Call for numerical application}

In this section, we present a methodology to calibrate the parameters of the forward-looking diffusion model, from accessible historical data sets.

\subsection{Historical dynamic}

The historical worldwide GDP growth rate is very well documented in the economic literature. It must be completed by an historical analysis of the global economic cost of climate change, as well as an estimation of the global transition effort made so far. This data set will allow to calibrate the cumulative quantities $\tilde{Y}_E^t$, $\tilde{Y}_P^t$ and $\tilde{Y}_T^t$ from $-t_0$. It also allows calculating the standard deviation of the hypothetical climate-free economic factor $\tilde{Y}_E^t$.

In the first version of the model (How Banks Can Save the Planet), we proposed a calibration of against IMF data points. The World Economic Outlook database gives access to the history of GDP at a country-granular level since 1980 \cite{imf}. World Bank also provides us with accurate GDP time series since 1961 \cite{worldbank}.

\subsection{Carbon intensity without transition}

The historical anthropogenic emission of carbon dioxide at year $t$ $CO_2^t$ is well documented in the climatic literature. By definition $-t_0$ is the year from
which GHG emissions have started to modify climate by exceeding the net-zero budget. We now introduce $-t_1$ as the year when the transition effort started. The period $[-t_0,-t_1]$ allows a definition of the net carbon intensity of the economic activity in the absence of transition effort:

\begin{equation}
I=(CO_2^{-t_1}-CO_2^{-t_0}) / (\ln{GDP^{-t_1}}-\ln{GDP^{-t_0}})
\end{equation}

The UNFCC provides a table of GHG emissions by country since 1990 \cite{unitednations} and the World Resources Institutes references all kinds of GHG emissions sources including PIK since 1850 and UNFCC data \cite{wri}.

\subsection{Transition efficiency}

Several economic organizations such as the OECD have estimated the expected economic cost $Y_{NZ}$ and duration $T_{NZ}$ to transition the global activity to net-zero carbon emission. We also note $R_{NZ}$ the average economic growth assumption made by the study. By taking the expected value of equation (2) we get the following relationship between the climate change intensity of the current economic activity $\tilde{\gamma}$ and the transition efficiency $\tilde{\alpha}$:

\begin{equation}
\tilde{\gamma}(\tilde{Y}_E^0- \tilde{Y}_P^0-\tilde{Y}_T^0+R_{NZ} T_{NZ})-\tilde{\alpha}Y_{NZ}=0
\end{equation}

We shall note $\tilde{\alpha}$ is constant, meaning it is assumed there is no diminishing or increasing efficiency of transition effort. It is possible to bypass this assumption at the cost of heavier formulas by making $\tilde{\alpha}$ a function of $t$. However, it should not call into question the computational implementation of our model, but only the formal calculations. Another possibility is to stress this parameter in order to assess the sensitivity of climate risk to the progress of green technologies. 

\subsection{Cost of climate change}

Several economic organizations have documented the distribution of the yearly economic cost of climate change by scenario of excess carbon concentration in the atmosphere, with long maturity $T$ (typically $30$ years). The OECD published in 2017 a study named Investing in Climate, Investing in Growth in which the cost of the transition is very well described \cite{oecd}. The climate scenarios proposed by the NGFS have prompted the ECB to calculate their respective associated GDP costs, both in terms of transition cost and climate change costs \cite{ecb}. It allows calibrating $a$ and $\tilde{p}$ in the equation describing the economic cost of climate change at time T: $\tilde{Y}_P^T-\tilde{Y}_P^{T-1}=a (CO_2^T-CO_2^{-t_0}) + \tilde{p} W_P^T$.

By substituting with the carbon intensity in the absence of transition efforts $I$ defined by (35), we get $\tilde{Y}_P^T-\tilde{Y}_P^{T-1}=a I (\tilde{Y}_E^T- \tilde{Y}_P^T-\tilde{Y}_T^T)-\tilde{\alpha}Y_T^T+ \tilde{p} W_P^T$. Hence, $\tilde{\gamma}=a I$ and we can deduct $\alpha$ from the equation (36). 

\subsection{Transition politics}

The transition politics parameters $\beta$ and $\theta$ result from a linear regression from $-t_1$ between the historical transition effort $(\tilde{Y_T^{t+1}}-\tilde{Y_T^{t}})$ at time $t$  and the historical cost of climate change $(\tilde{Y_P^{t}}-\tilde{Y_P^{t-1}})$ at time $t-1$.

\newpage

\printbibliography
\end{document}